\documentclass[5p,twocolumn,times,number]{elsarticle}

\usepackage{graphicx}
\usepackage[squaren]{SIunits}

\newcommand{\textroman}[1]{\mbox{\rm #1}}
\newcommand{\gfrac}[2]{\displaystyle\frac{#1}{#2}}

 \newcommand{\Aeff}{A_{\textroman{eff}}}

\journal{Nucl.\ Instrum.\ Meth.\ A }

\begin{document}

\begin{frontmatter}

\title{TPC in $\gamma$-ray astronomy above pair-creation threshold }

\author[add1]{D.~Bernard\corref{cor}}
\ead{denis.bernard at in2p3.fr}

\address[add1]{LLR, Ecole Polytechnique, CNRS/IN2P3,
91128 Palaiseau, France}


\begin{abstract}
We examine the performance of a TPC as a $\gamma$-ray telescope
above the pair-creation threshold.
The contributions to the photon angular resolution are studied and
their dependence on energy is obtained.
The effective area per detector unit mass for such a thin detector is the conversion mass
attenuation coefficient.
The differential sensitivity for the detection of a point-like source 
is then derived.
Finally, the measurement of track momentum from deflections due to
multiple scattering is optimized.

These analytical results are exemplified numerically for a few sets of
detector parameters.
TPCs show an impressive improvement in sensitivity with respect to
existing pair-creation-based telescopes in the [MeV - GeV] energy
range, even with the modest detector parameters of this study.
In addition, gas TPCs allow an improvement in angular resolution of
about one order of magnitude.
\end{abstract}


\end{frontmatter}

\section{Introduction}

High-energy
astrophysics focuses on the study of complex objects such as active galactic
nuclei (AGN), pulsars and gamma-ray bursts (GRB), the spectral energy
distributions (SEDs) of which comprise contributions from a number of
processes \cite{Ghisellini:2012hs}, including synchrotron emission,
synchrotron self-Compton scattering (SSC),
the inverse scattering of external thermal photons from 
 the accretion disk (ESC), and $\pi^0$ decays.
These studies suffer from the existence of gaps in the sensitivity of
the available instruments, such as that between the
Compton-scattering-based telescopes, that are mostly sensitive below
1 MeV, and the pair-based telescopes, that are mostly sensitive above
100 MeV.
Pair-based telescopes also suffer from a poor angular
resolution at low energy, making the analysis of dense regions of the sky 
difficult.

In this paper we present a characterization of the use of a thin
detector as an active target for cosmic $\gamma$-ray detection in the
pair-conversion regime, with the aim of improving the angular
resolution and the sensitivity at low energy.
The past missions and the presently operated Fermi/LAT and AGILE use
thick detectors for which the photon conversion probability is close to
unity and the effective area $\Aeff$ is the product of the geometrical
area of the detector $A$ and of the reconstruction efficiency
$\epsilon_r$: $\Aeff = A \times \epsilon_r$.
In a thin detector for which the conversion probability is small, the
effective area becomes proportional to the pair conversion mass
attenuation coefficient $H$ and to the detector mass $M$, $\Aeff = H
\times M \times \epsilon_r$.
In a thick detector, the optimization of the aspect ratio
height/surface is therefore critical, while for a thin detector it does
not affect the effective area to first order.
Note that in a thick detector, the various possible conversion
processes are in competition with each other, and Compton scattering
prior to- or in place of- pair conversion would become a problem at
the lowest energy if low $Z$ material were considered, an effect
strongly reduced for a thin detector.
A sample of previous works on the use of  gas detectors for the
detection of cosmic $\gamma$ rays above the pair-creation threshold,
sometimes in relation with a Compton telescope, can be found in Refs.
\cite{Hartman,Hunter:2001,Ueno:2011}.

\section{Test detector parameters}

We use a time projection chamber (TPC) as a test model of a thin
detector. A TPC is a chunk of homogeneous matter, located inside a
(say) uniform electric field $\vec{E}$
(\cite{Nygren:1978}, and a recent review \cite{Attie:2009zz}).
Charged tracks crossing the TPC ionize atoms/molecules on their path,
after which the ionization electrons drift to the anode plane where
(in a gas TPC) they are amplified, and collected.
The collecting plane can be segmented so as to provide two coordinates
$x, y$ of the location of the ionization.
The third coordinate $z$ is provided by the drift time $t$, 
$z = v_{d} \times t$.
The drift velocity $v_d$ ranges from 
$ {\cal O}(\milli\meter/\micro\second)$ at saturation for liquids and
 solids,
to $v_d = {\cal O}(\centi\meter/\micro\second)$ for gases when used
with an appropriate quencher, a multi-atomic molecule  on which 
drifting electrons collide and are cooled down.
The size of the electron ``cloud'' increases during the drift due to
diffusion, which places a limit to the useable drift length.
The typical value of the diffusion coefficient is 
 $ {\cal O} (100 - 200 \, \micro\meter/\sqrt{\centi\meter})$
for the three phases considered here.

Note that 
amplification of the signal is easily performed in a gas, but not
in a liquid.
The smallness of the collected signal $\approx 4000 \, e^-/\milli\meter$
in liquid argon (lAr) and the limitation of the electrical power
available in space for the digitization electronics would make precision
tracking difficult.
To alleviate this limitation, double-phase systems have been
developed: the electrons are extracted from a liquid TPC into a gas
in which they are amplified efficiently
\cite{Bondar:2005wx,Badertscher:2010zg}.
But for a use in space, the instability of the gas/liquid interface
would be a problem.
A gas/solid system might be considered such as solid argon and a
neon-based gas.
The density and electron parameters of noble-gas solid and liquid are
close enough that we neglect the difference and we use the liquid
as representing the two dense phases\footnote{Note that liquid neon
 does not allow electron to drift, but solid neon does
 \cite{Brisson:1983qb}.}.

These analytical results are exemplified numerically for a few sets of
detector parameters :
three noble gases: neon, argon and xenon; 
three ``densities'': gas at 1 and 10 bar, and liquid.
We chose on purpose rather modest detector parameters, a detector mass
$M= 10\, \kilo\gram$ and a tracking length of $L = 30 \, \centi\meter$ 
for gas, $M= 100\, \kilo\gram$ and $L = 10 \, \centi\meter$ for liquid,
except in the case where the radiation length $X_0$ would be shorter than $L$: in
that case the propagation of an electron would be affected by
radiation after a path length of the order of $X_0$, and we use
$L=X_0$.
The tracking length $L$ is obviously related to the detector
thickness, but they can be different, as in the case of liquid xenon.
The sensitive mass for a gas TPC is chosen to be smaller due to the volume
limitation of space missions.

We use typical values for the longitudinal sampling of the TPC, $l = 1
\milli\meter$, and of the point space resolution, $\sigma =
0.1\milli\meter$.

\section{Angular resolution}

We consider a photon with energy $E$ converting to an
electron-positron pair in the field of a charged particle of the
detector. 
The conversion is said to be
``nuclear'' in the case of an ion, $\gamma Z \rightarrow e^+e^- Z$, and
``triplet'' in the case of an electron $\gamma e^- \rightarrow e^+e^- e^-$.
From momentum conservation:
\begin{equation}
 \vec{p_{\gamma}} = \vec{p_{e^+}} +\vec{p_{e^-}} + \vec{q},
 \label{eq:momentum:conservation}
\end{equation}
we can identify several contributions to the photon angular resolution:
(1) the single track angular resolution;
(2) the fact that, in the case of nuclear conversion, the recoil
momentum $q$ of the ion, of the order of $1 \, \mega\electronvolt/c$
produces a path length that is too small to allow a measurement of $q$;
(3) the resolution of the norm of the momentum, named here 
the momentum resolution, for each of the hard tracks.
We examine these three contributions in the following.

\subsection{Single track angular resolution}

The basis of the understanding of tracking 
in the presence of multiple scattering was settled by 
Gluckstern \cite{Gluckstern}.
We make use of the similar results derived by Innes for optimal fits
\cite{Innes:1992ge}.
We use an approximation of the multiple scattering angle $\theta_0$
undergone by a particle of momentum $p$ crossing a slab of matter with
thickness $x$ \cite{PDG}: 
\begin{equation}
\theta_0 = \gfrac{p_0}{\beta cp} 
\sqrt{\gfrac{x}{X_0}},
 \label{eq:multiple:scattering:base}
\end{equation}
where 
$p_0 = 13.6 \, \mega\electronvolt/c$ and 
we have neglected the negative logarithmic correction factor.
At high momentum, multiple scattering can be neglected and the detector
resolution dominates
\cite{Innes:1992ge} :
\begin{equation}
\sigma_{\theta tH} \approx \gfrac{8 \sigma}{L}\sqrt{3/(N+5)},
\label{eq:bb:noscat}
\end{equation}
 where $N$ is the number of samplings, $N = L/l$.
In the expressions of $\sigma_{\theta}$, the subscript $t$ refers to a
track and $H$ and $L$ to the high and low track momentum,
respectively.
$\sigma_{\theta tH}$ is independent of the track momentum and improves like 
 $L^{-3/2}$ at given sampling $l$.
On the contrary, multiple scattering dominates at low momentum; from
the results of Ref. \cite{Innes:1992ge} and using the approximation
of eq. (\ref{eq:multiple:scattering:base}), we obtain:
\begin{equation}
\sigma_{\theta tL} 
 \approx 
(2 \sigma)^{1/4} l^{1/8} X_0^{-3/8} (p/p_0)^{-3/4}.
 \label{eq:Vbb:p:2}
\end{equation}

The expression does not depend on the tracking length.
This is because the end of the track
 contributes no information
about the particle's direction at the detector entrance once
there has been sufficient scattering \cite{Innes:1992ge}.
The $p^{-3/4}$  dependence on momentum obtained here is comparable to the 
 $E^{-0.78}$ dependence on energy parametrized by the Fermi Collaboration \cite{Fermi:1206.1896}.
The momentum limit between the two regimes is at
$p=p_{lim}$, with:
\begin{equation}
p_{lim} = p_0 \times a_p
\frac{L^2}{\sigma \sqrt{X_0 l}}, 
 \label{eq:plim}
\end{equation}
where $a_p = (2/8^4 \times 9)^{1/3}\approx 0.038$.

\subsubsection{From tracks to photon}

Combining the measurements of the directions of the electron and of
the positron in the small-angle approximation, the direction of the
reconstructed photon (with respect to the (unknown) true incoming
direction) is $\theta_\gamma = r \theta_{x+} +(1-r) \theta_{x-}$,
where $r$ is the fraction of the energy that is carried away by the
positron and $\theta_{x+}$ and $\theta_{x-}$ are the angles of the
positron and of the electron, respectively.
We then compute the photon angular resolution from the track 
 angular resolution, eqs. (\ref{eq:bb:noscat}),(\ref{eq:Vbb:p:2}).

\begin{figure} [th]
 \begin{center}
 \includegraphics[width=0.64\linewidth]{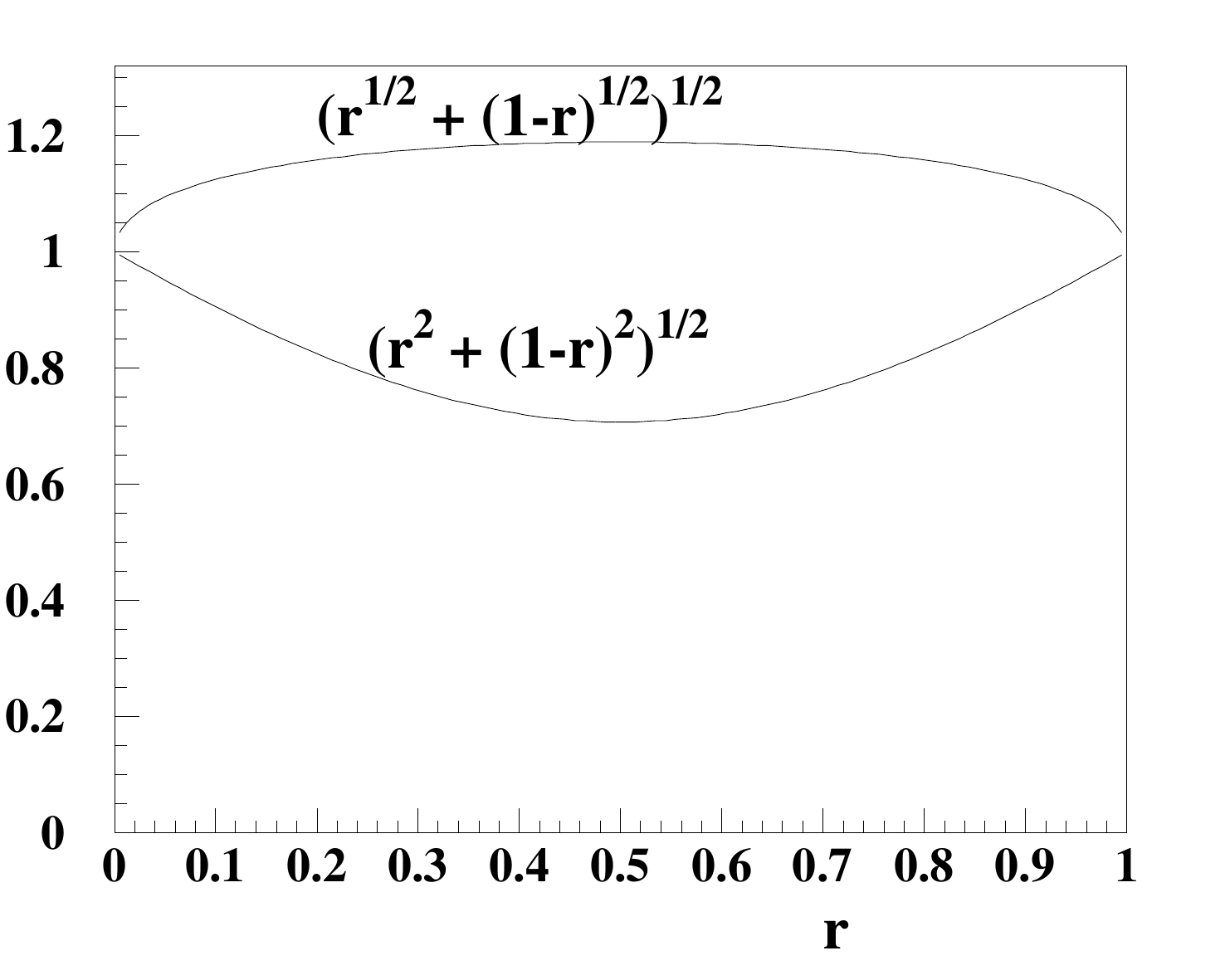}
 \caption{\label{fig:yuy}
Track-to-photon angular resolution factors as a function of the fraction $r$ of the photon energy carried away by the positron.
}
 \end{center}
\end{figure}

In the multiple scattering dominated regime we obtain
$ \sigma_{\theta \gamma} = \sigma_{\theta t} \sqrt{\sqrt{r}+ \sqrt{1-r}} $
and in the 
high energy regime: 
$ \sigma_{\theta \gamma} = \sigma_{\theta t} \sqrt{r^2+ (1-r)^2} $,
in the expressions of which the track momentum $p$ is replaced by the
photon energy $E$.
The photon angular resolution is therefore obtained from the single
track angular resolution by applying a correction factor that is close
to unity (Fig.\ref{fig:yuy}) and that we neglect in the following.

\begin{figure} [th]
 \begin{center} 
 \includegraphics[width=0.48\linewidth]{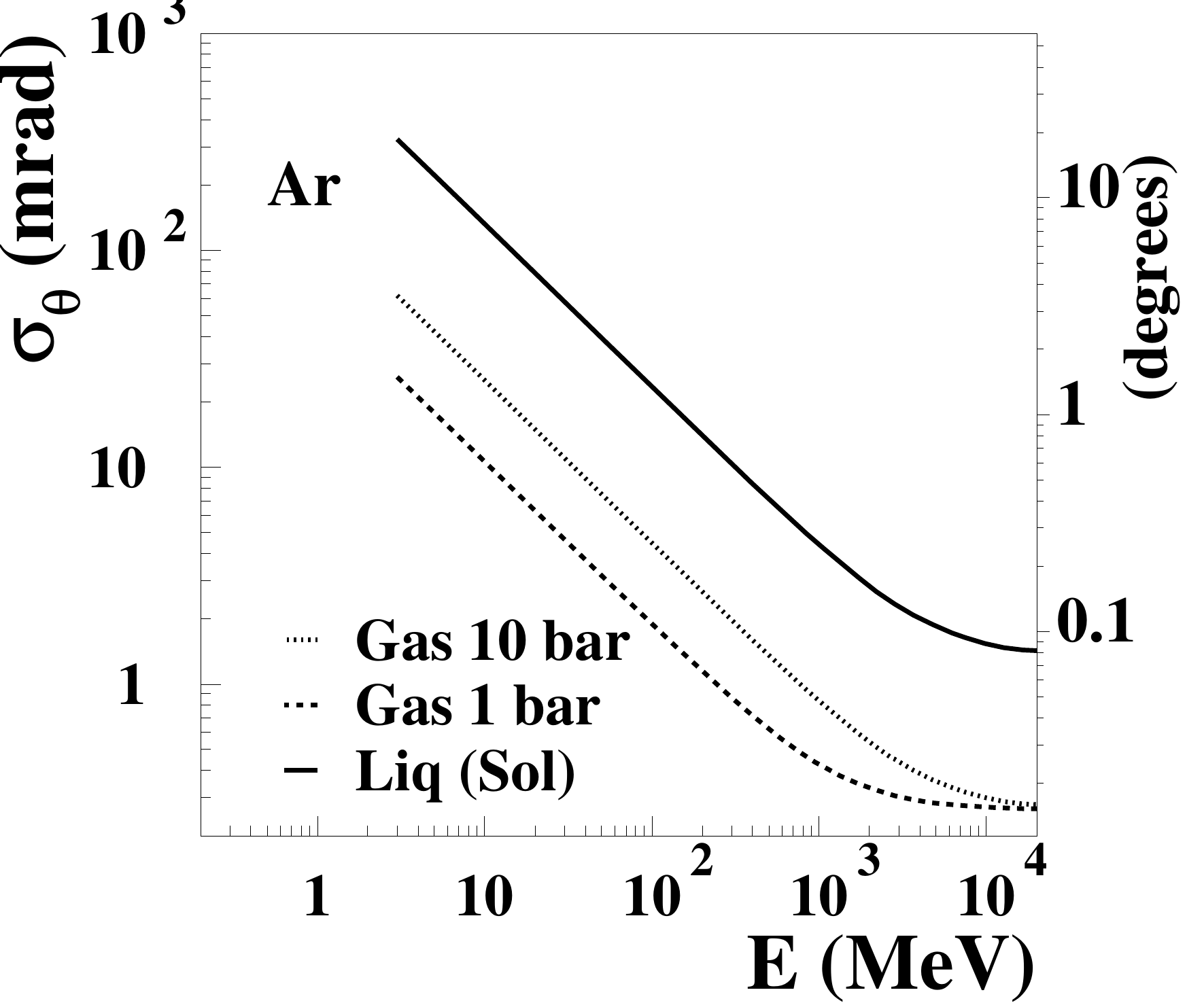}
 \includegraphics[width=0.48\linewidth]{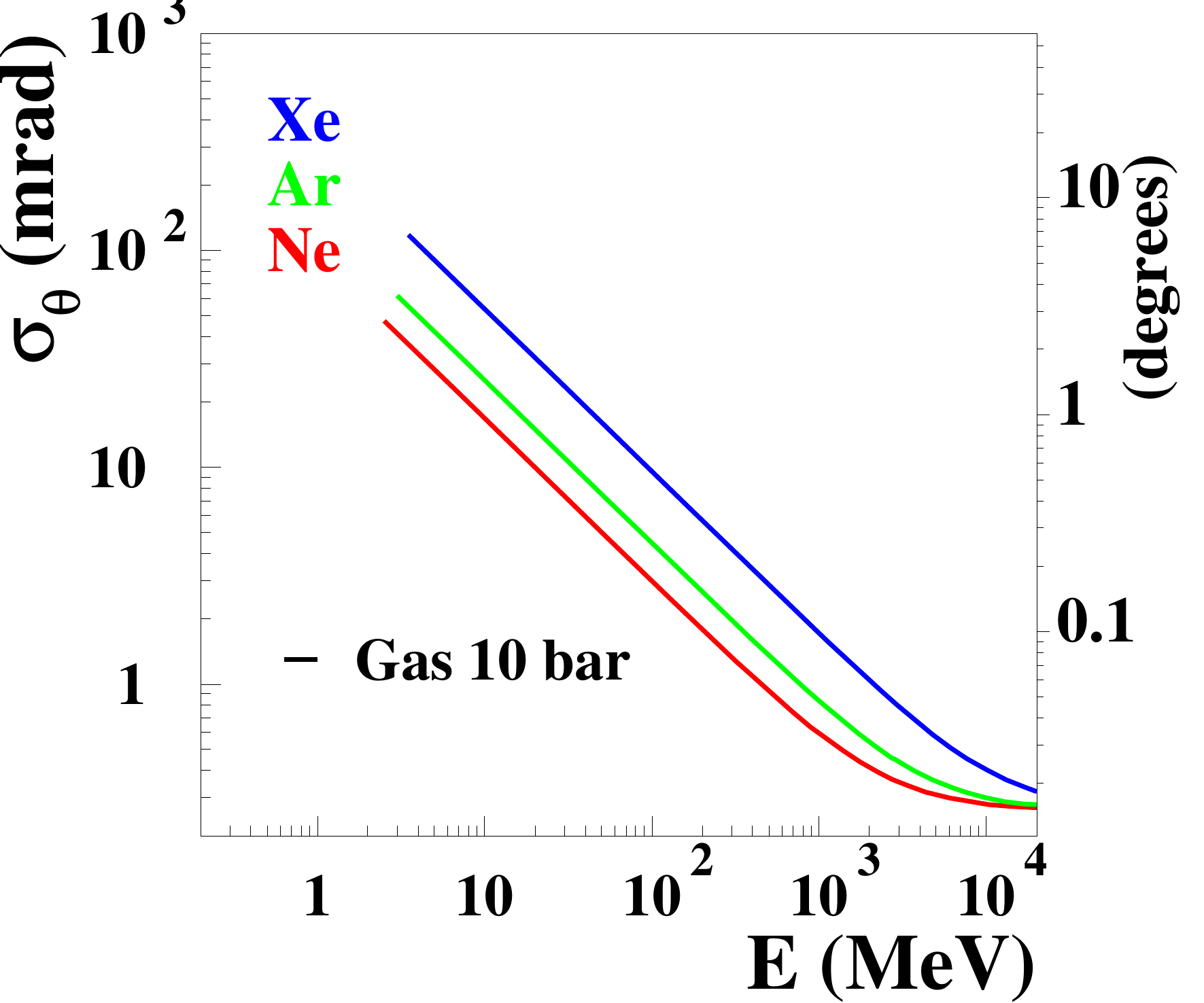}
\caption{\label{fig:res:ang:vars}
(Color online).
Dependence of the RMS photon angular resolution on photon energy, 
in argon for various densities (left) and in 10 b gas 
for various gases (right).
}
\end{center}
\end{figure}
Assuming a quadratic sum of $\sigma_{\theta tH} $ and of
$\sigma_{\theta tL} $ in the intermediate energy range, we can examine
the energy dependence of the photon angular resolution due to single
track angular resolution (Fig. \ref{fig:res:ang:vars}).
Less dense TPCs have a better resolution.
The difference in the high-energy asymptotic values, between gas and
liquid, is due to the shorter tracking length used here, 10~cm vs 30~cm.

\subsubsection{Recoiling ion}

\begin{figure} [th]
 \begin{center}
 \includegraphics[width=0.64\linewidth]{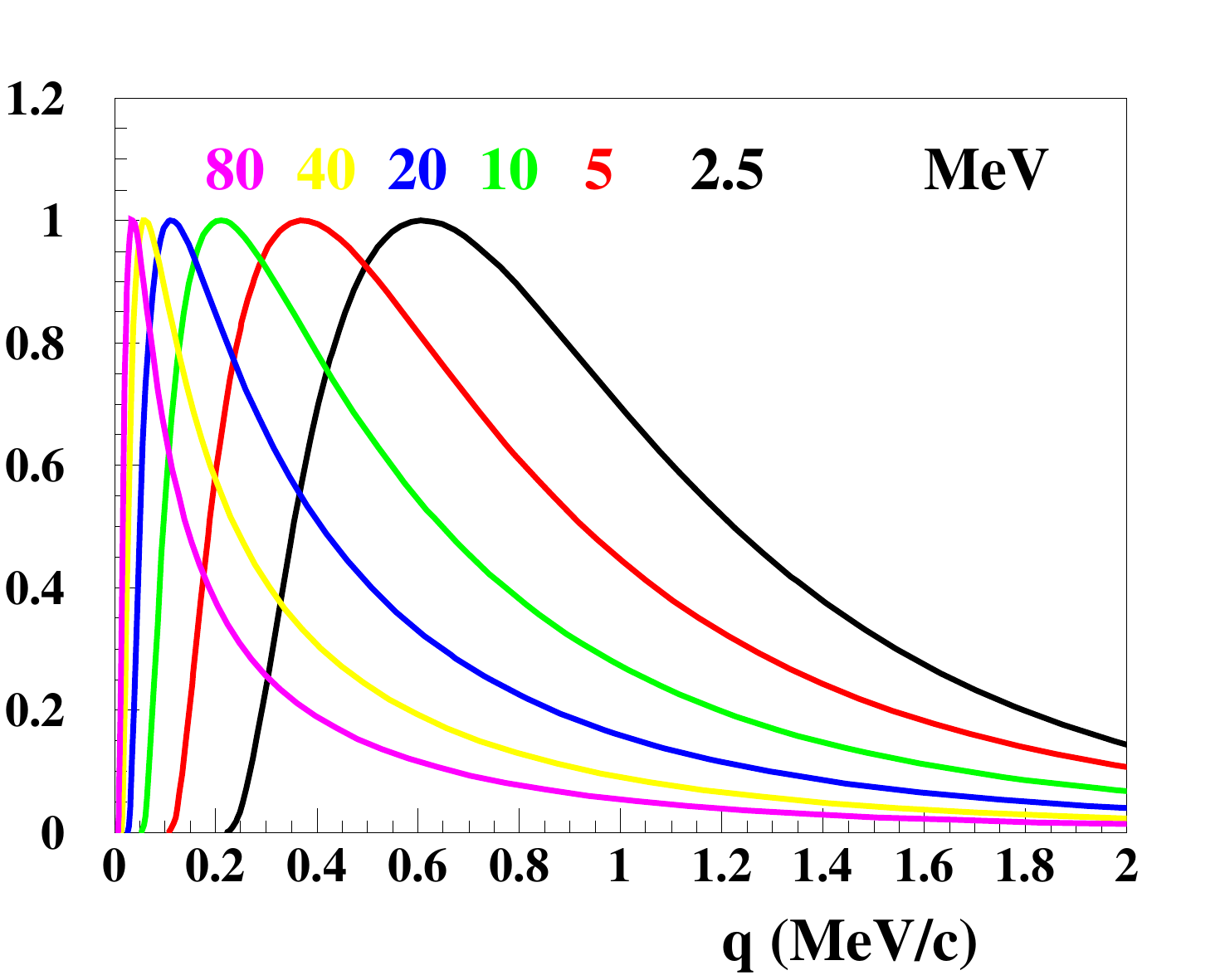}
 \caption{\label{fig:momentum:trace}
(Color online).
Distribution of the recoil momentum for several incoming photon
energies, from Ref. \cite{Jost:1950zz}, in the approximation of no
screening, normalized to unity at their maximum.}
 \end{center}
\end{figure}

In the case of nuclear conversion, the recoil momentum goes
undetected, unless a very low pressure TPC is used.
The recoil is almost transverse to the photon direction, and the
contribution to the photon angular resolution is therefore $\approx q/E$.
The $q$ distribution for nuclear conversion has been obtained by Jost
et al. \cite{Jost:1950zz} (\footnote{With Borsellino's correction \cite{Borsellino1953} applied.})
by the integration of the 5D Bethe-Heitler differential cross section
\cite{Heitler1954}.
It has a very wide spectrum that extends from $2 m^2/E$ to $E$
 at high energy, where $m$ is the electron mass.
The spectrum  peaks at a value of the recoil momentum $q_M$ that 
 decreases with $E$ (Fig. \ref{fig:momentum:trace}).

\begin{figure} [th]
 \begin{center} 
\includegraphics[width=0.48\linewidth]{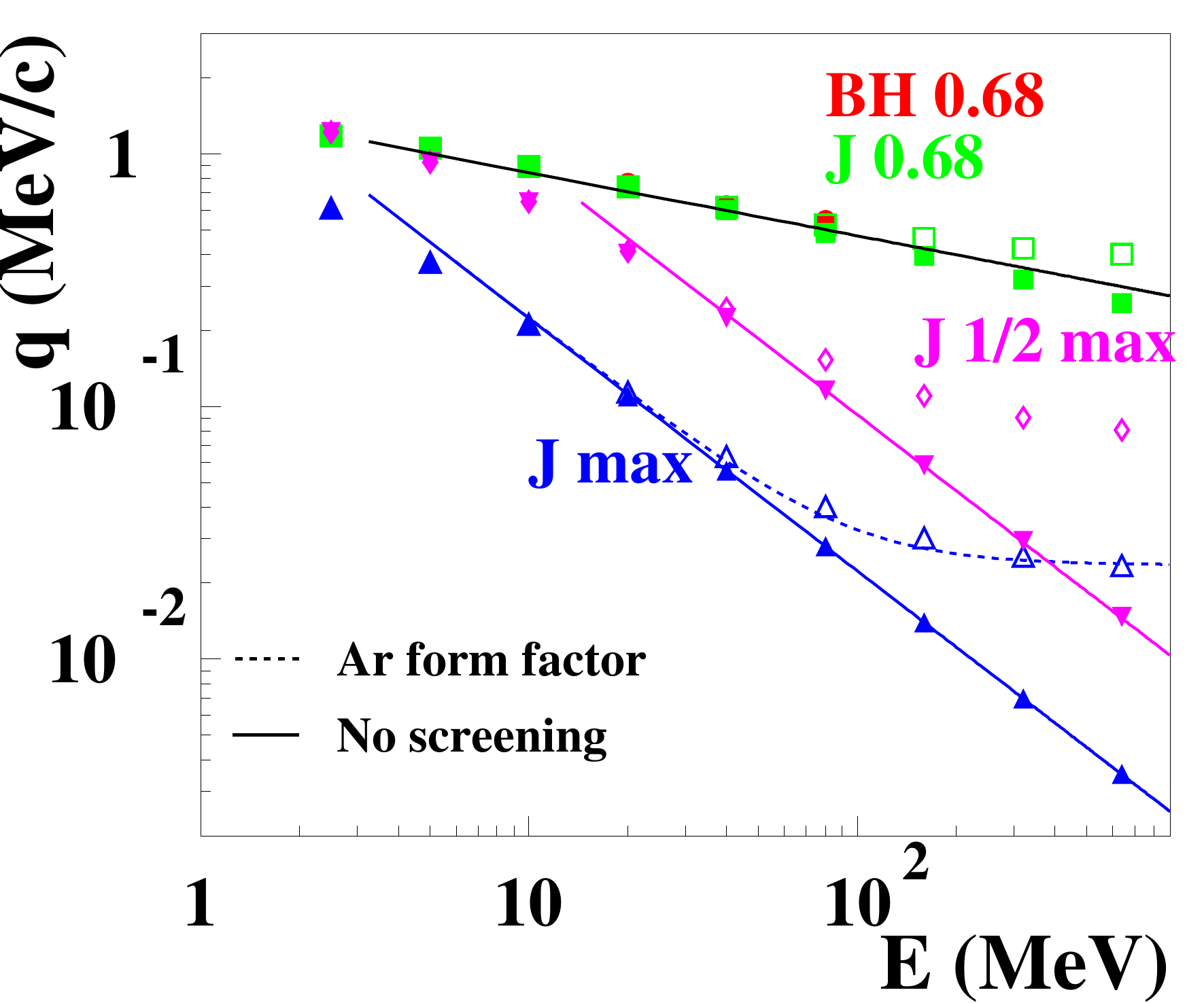}
\includegraphics[width=0.48\linewidth]{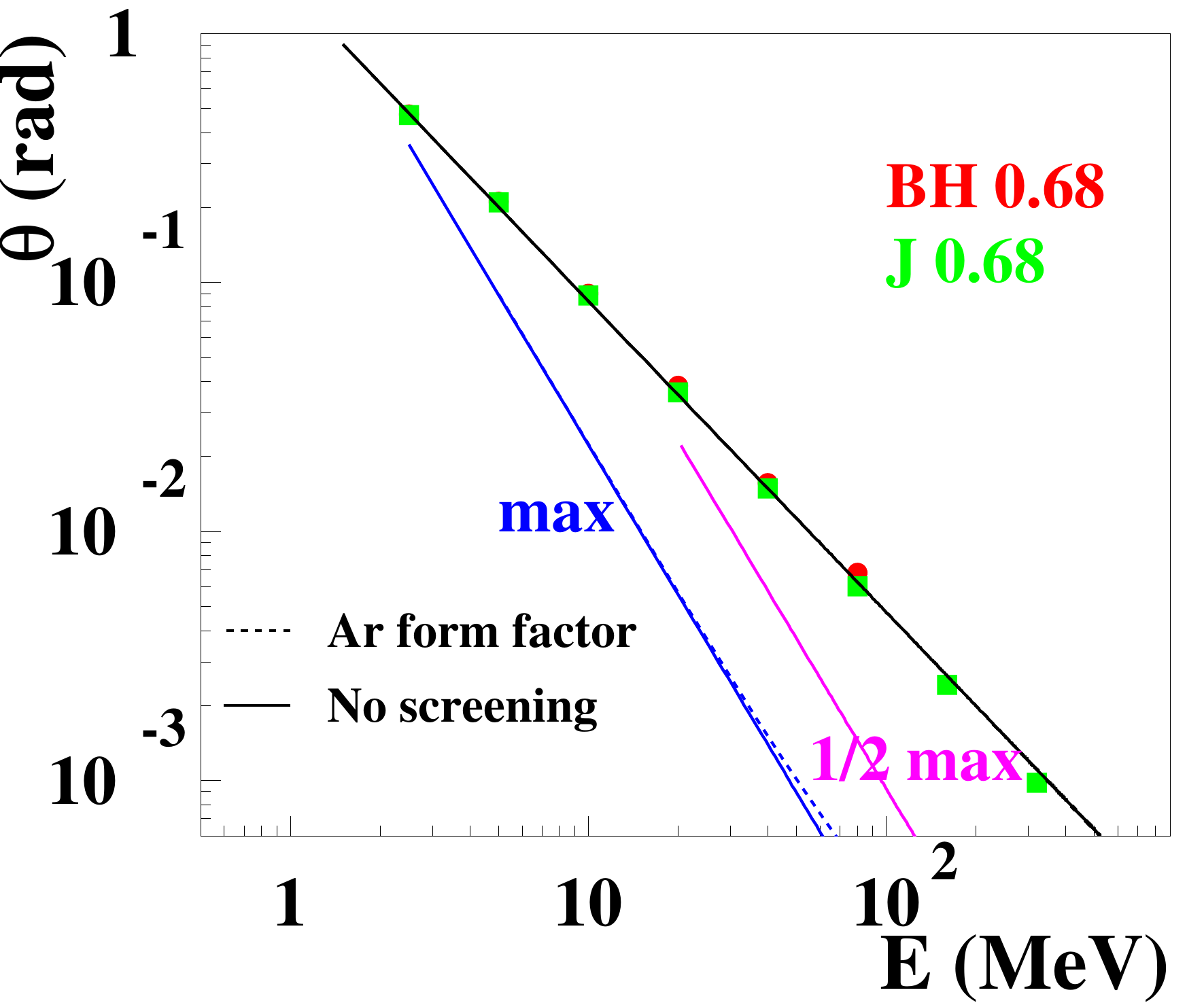}
\caption{\label{fig:moment:recul:trace:FF}
(Color online).
Left: 
68 \% ``containment'' value $q_{68}$ of the recoil momentum $q$, as
a function of incoming photon energy $E$.
'J' stands for the $q$
distribution by Jost et al. (squares)
\cite{Jost:1950zz}, while 'BH' stands for the 5D Bethe-Heitler
differential cross-section (bullets) \cite{Heitler1954}.
Values $q_M $ of $q$ at maximum (up triangles) and at half maximum $q_{1/2}$
(down triangles).
The solid symbols corresponds to the absence of screening, and the
open symbols to screening parametrized by the Mott form factor \cite{Mott:1934}.
Right: 
68 \% ``containment'' space angle induced by the non-observation of the recoiling nucleus.
The upper line corresponds to $\theta = 1.5 \, \radian [E / 1\, \mega\electronvolt]^{-5/4}$,
the lower lines to $q_{1/2}/E$ and $q_M /E$. 
}
\end{center}
\end{figure}

The 68 \% containment value $q_{68}$ of $q$ is of particular interest as the
point spread function of $\gamma$ telescopes is often expressed in that way.
The estimations of $q_{68}$ from the 1D $q$ distribution
\cite{Jost:1950zz} and from the 5D Bethe-Heitler differential cross
section \cite{Heitler1954} are found to be compatible with each other which
constitutes a cross check (Fig. \ref{fig:moment:recul:trace:FF}).
Also of interest is the momentum at half-maximum $q_{1/2}$, on the
decreasing slope above the maximum.

Under the approximation of no screening of the field of the nucleus by
the surrounding electrons, we find $q_M$ and $q_{1/2}$ to vary like 
$\approx 2.24 \, \mega\electronvolt / (c E)$
and 
$\approx 9.28 \, \mega\electronvolt / (c E)$, respectively, 
while  $q_{68}$ has a much milder dependence on $E$,
as
$q_{68} \approx (1.5 \, \mega\electronvolt/c) E^{-1/4}$.
Therefore, at high energy, a low-$q$ core develops that is narrower with
increasing $E$, but that contains a fraction of events which
decreases with $E$.

\begin{table}[h]
 \begin{center} \small
\begin{tabular}{ll|cc|cc}
\hline
& & \multicolumn{2}{c|}{No screening} & \multicolumn{2}{c}{Screening} \\
& & \multicolumn{2}{c|}{(low energy)} & \multicolumn{2}{c}{(high energy)} \\
\hline
& & $q$ & $\theta$ & $q$ & $\theta$ \\
\hline
\noalign{\vskip1pt}
 $q_{68}$ & 68 \% containment & $\propto E^{-1/4}$ & $\propto E^{-5/4}$ & constant & $\propto 1/E$ \\
 $q_M$ & at maximum & $\propto 1/E$ & $\propto 1/E^2$ & constant & $\propto 1/E$ \\
 $q_{1/2}$ & at half max & $\propto 1/E$ & $\propto 1/E^2$ & constant & $\propto 1/E$ 
\end{tabular}

\caption{Dependence on photon energy of $q_{68}$, $q_{M}$, and $q_{1/2}$ and related angular shifts from $\theta \approx q/E$. \label{tab:q:theta}} 
\end{center} 
\end{table}

The dependence of $q_M$, $q_{1/2}$ and $q_{68}$ on photon energy are represented in
Fig. \ref{fig:moment:recul:trace:FF}, and tabulated in Table \ref{tab:q:theta}.
Closed symbols refer to the absence of screening. The screening of the
field of the nucleus by the surrounding electrons can be described by
multiplying the $q$ distribution \cite{Jost:1950zz} by $(1 - F(q))^2$,
where $F(q)$ is the Mott atomic form factor $F(q) = \frac{1}{1 +
 (q/q_0)^2}$, with $q_0 = 111 Z^{-1/3} \times m$ \cite{Mott:1934}.
When screening is taken into account, the 68\% containment value
saturates at high energy to a value that has a mild $Z$ dependence,
and that is close to $0.4 \, \mega\electronvolt/c$, while the
distribution maximum $q_M$ saturates to a much smaller value close to
$2 \times q_0$.
The induced photon angle shift is obtained simply by $\theta = q/E$ (Fig. 
 \ref{fig:moment:recul:trace:FF} right).

\subsection{Contribution of track momentum resolution}

The contribution of the track momentum resolution $\Delta p$ in the
reconstruction of the photon momentum is estimated using a generator
of the 5D Bethe-Heitler \cite{Heitler1954} differential cross section,
to which a momentum spread assumed to be of 10 \% is added.
\begin{figure} [th]
 \begin{center} 
\includegraphics[width=0.64\linewidth]{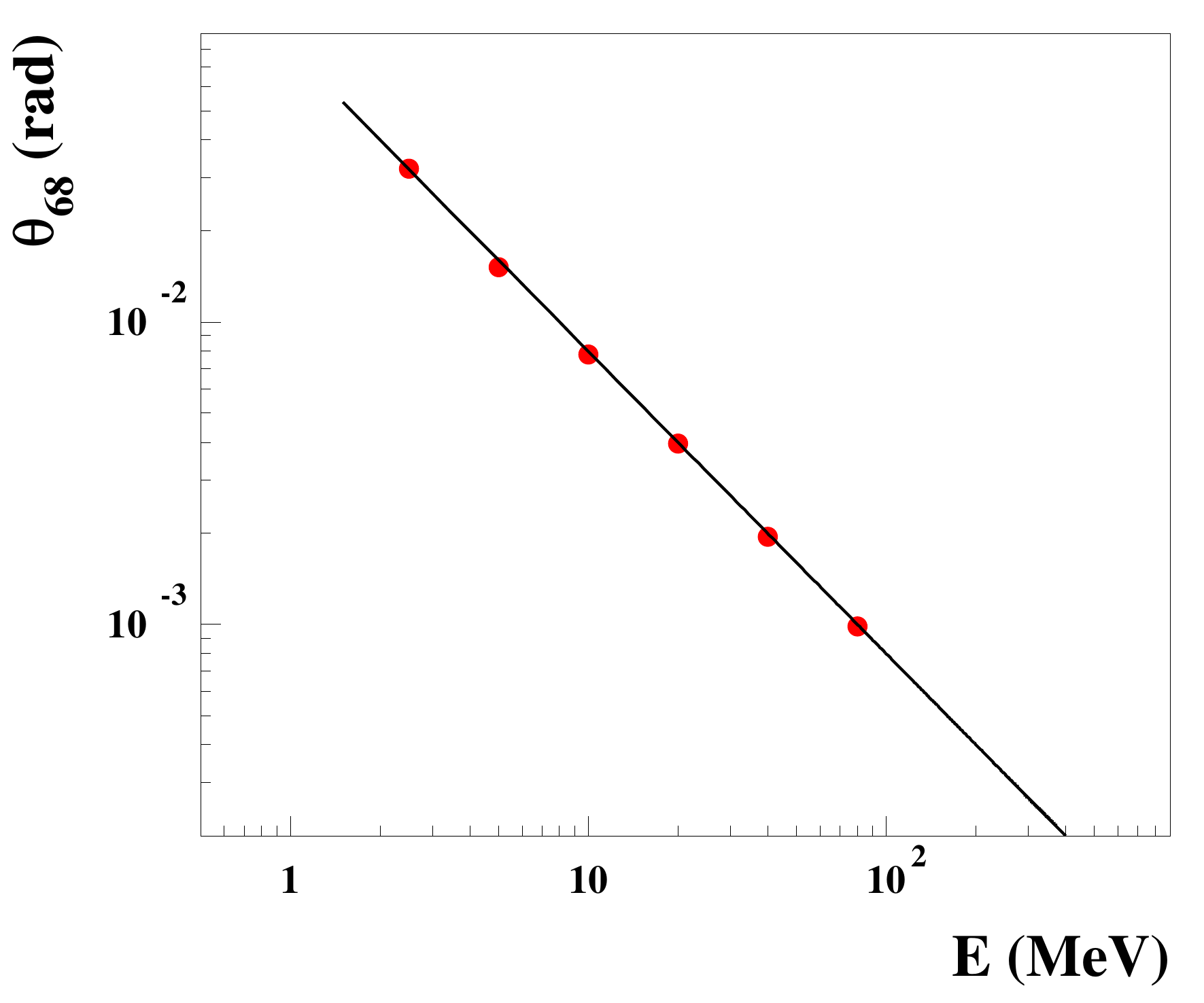}
\caption{\label{fig:ang:res:momentum:blur}
(Color online).
68\% containment angle shift due to a track momentum resolution of $ \Delta p/p = 10 \%$, from a generation of the 5D Bethe-Heitler 
\cite{Heitler1954}
differential cross section.}
\end{center}
\end{figure}

The dependence is found to be 
$\theta \approx (\Delta p/p) 0.8 \, \radian/E$
(Fig. \ref{fig:ang:res:momentum:blur}).
This $1/E$ dependence was expected given the $1/E$ dependence of the
opening angle \cite{Olsen:1963zz}.

\subsection{Angular resolution: a Summary}

\begin{figure} [th]
 \begin{center} 
 \includegraphics[width=0.8\linewidth]{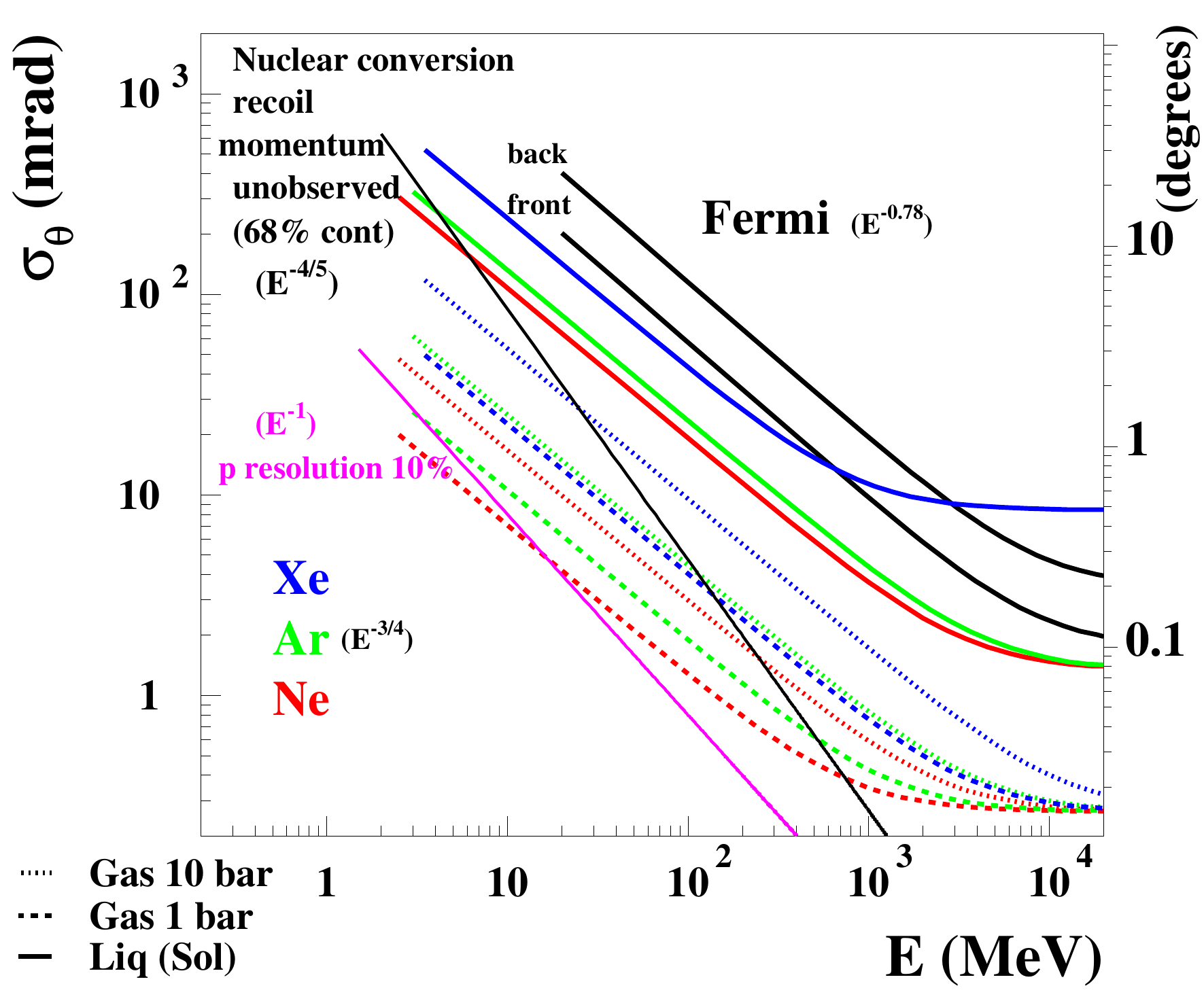}
\caption{\label{fig:ang:res:summary}
(Color online).
Various contributions to the photon angular resolution.
}
\end{center}
\end{figure}

The results obtained so far are summarized in
Fig. \ref{fig:ang:res:summary} (we have neglected the small difference
between the RMS and the 68\% containment values).

\begin{itemize}
\item 
The single track angular resolution dominates;
\item 
The missing
momentum recoil contributes significantly at low energy,
especially for low density TPC;
\item 
The effect of a single track momentum resolution at a level of 
10~\% as assumed here is negligible.
\end{itemize}

A ten-fold improvement of the angular resolution with respect to the
performance of the Fermi/LAT \cite{Fermi:1206.1896} is within reach
with a gas TPC. 
A liquid xenon TPC shows a poor performance here, in contrast with its
use as a Compton telescope\cite{Aprile:2008ft}.

\section{Effective area}

\begin{figure} [th]
 \begin{center} 
 \includegraphics[width=0.84\linewidth]{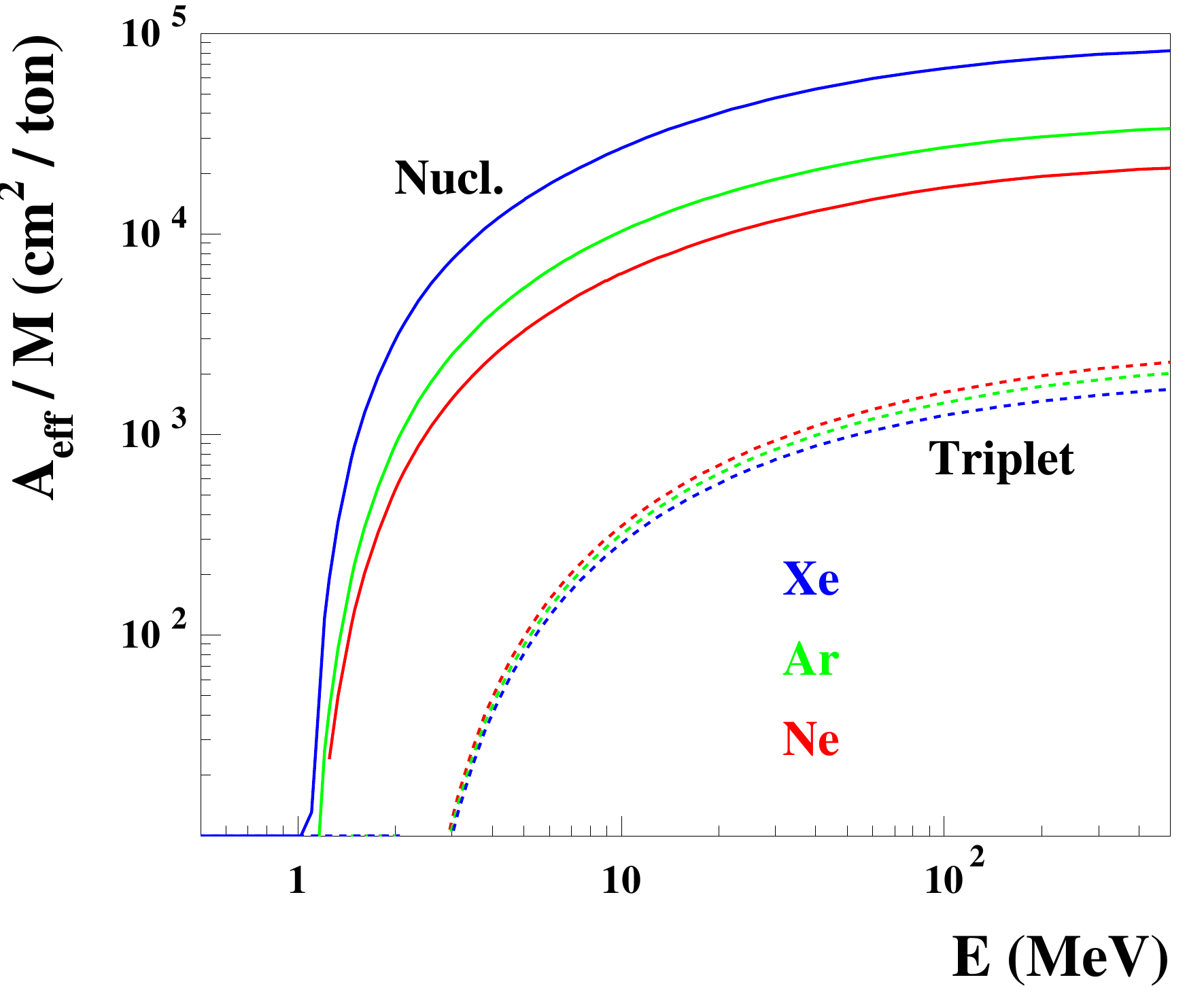}
\caption{\label{fig:Aeff}
(Color online).
Dependence on energy  of the effective area per unit mass assuming 100\%
reconstruction efficiency \cite{NIST}. }
\end{center}
\end{figure}

Assuming 100\% reconstruction efficiency, which is reasonable for a
TPC, 
 we obtain the dependence on energy of $\Aeff $ (Fig. \ref{fig:Aeff}) 
from the tabulation of the mass attenuation coefficient in Ref. \cite{NIST}.

In the case of triplet conversion, the effective area is almost
independent of $Z$ which is not surprising, as the number of electrons
per unit mass is almost a constant of Nature.

\section{Differential sensitivity}
\label{sec:sens}

From the above we compute the sensitivity to the detection
of a faint high-latitude point-like source.
We mimic the computation of the Fermi/LAT sensitivity 
\cite{Fermi:1206.1896} with a simpler method.

We use four bins per energy decade 
($\Delta E = 10^{-1/4} E \approx 0.78 E$),
a $n=5$ standard deviation significance, 
a $T = 3 \, \textroman{year}$ observation time, 
an $\eta = 17 \%$ exposure fraction, 
and a minimum signal event number $S\ge 10$.
The sensitivity is computed inside a $\sigma_\theta$ 68\%-containment
angle, and ``against'' the extragalactic $\gamma$-ray background
\cite{Abdo:2010nz}: we therefore compare our results to
 Fermi's differential sensitivity at $90^\circ$ galactic latitude
\cite{Fermi:1206.1896}.

\subsection{Gaussian statistics}

Using Gaussian statistics, $n$ is simply obtained from the numbers $S$
and $B$ of signal and background events, respectively, from $S = n
\sqrt{B}$.
We compute these numbers of events as :
\begin{equation}
S = T \times \eta \times \Aeff (E) \times I_0(E) \times \epsilon \times \Delta E
\end{equation}

\begin{equation}
B = T \times \eta \times \Aeff (E) \times \pi \sigma_{\theta}^2 \int F_B(E) dE
\end{equation}

where $F_B$ is the background flux, $I_0$ the signal intensity, and 
$\epsilon = 0.68$, the efficiency of the angle cut.
The sensitivity $s$ expressed as the minimum detectable signal
intensity, multiplied by $E^2$, that is $s = E^2 \times I_0$, is
therefore:
\begin{equation}
s = E^2 \times I_0 = \frac{n \times E^2} {\epsilon \Delta E} 
\sqrt{ \frac{ \pi \sigma_{\theta}^2 \int F_B(E) dE}{T \times \eta \times \Aeff }}
\label{eq:s:gaussian}
\end{equation}

The factors that depend on $Z$ are $\sigma_\theta \propto X_0^{-3/8} $ and
$\Aeff \propto H$ that is, asymptotically, $\Aeff \propto 1/X_0$.
These dependences cancel partially in the expression of $s \propto
\sigma_\theta / \sqrt{\Aeff }$, so that $s \propto X_0^{1/8} $: the $Z$
dependence of the sensitivity as described by eq. (\ref{eq:s:gaussian})
is extremely small.

\begin{figure*} [th]
\begin{center} 
\includegraphics[width=0.48\linewidth]{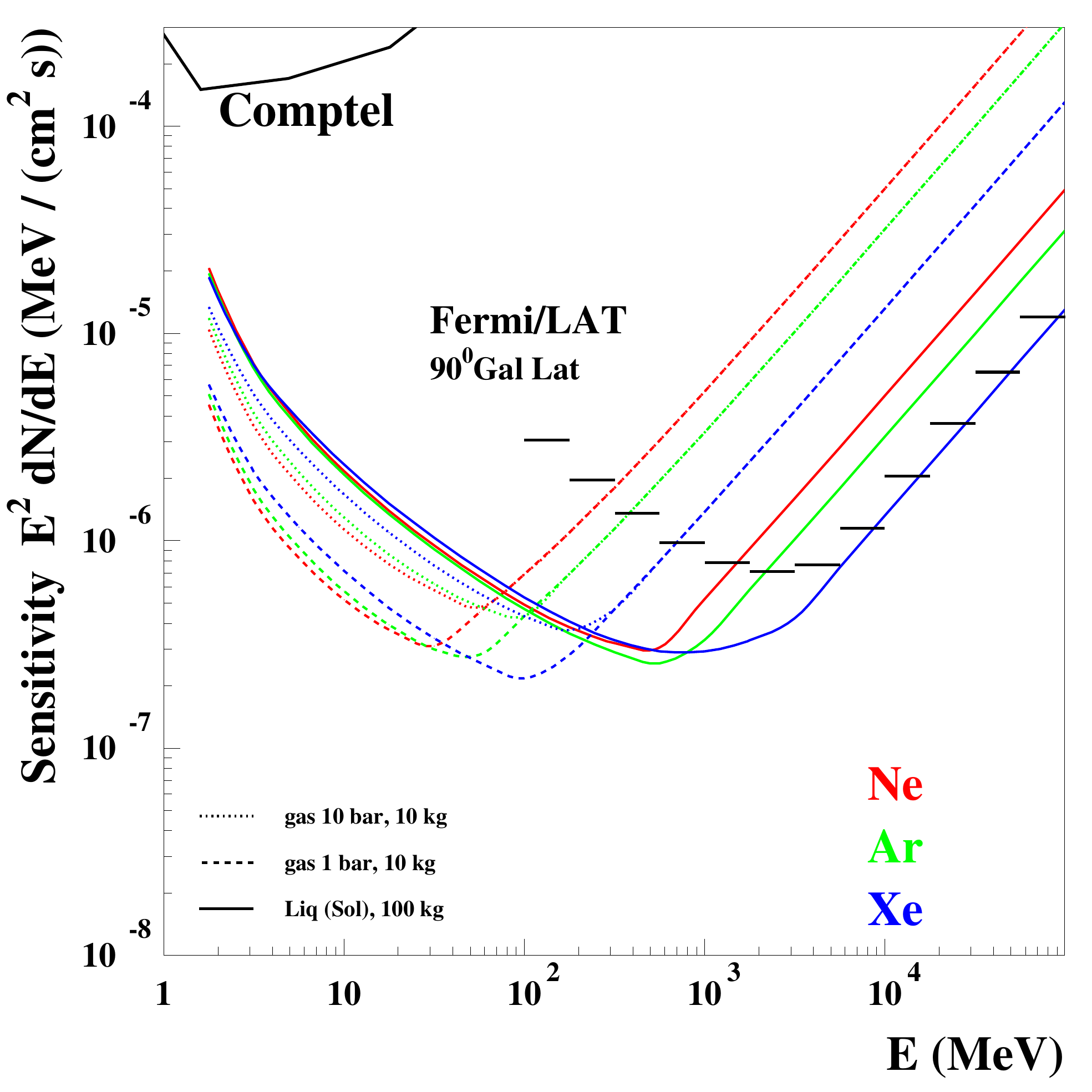}
\includegraphics[width=0.48\linewidth]{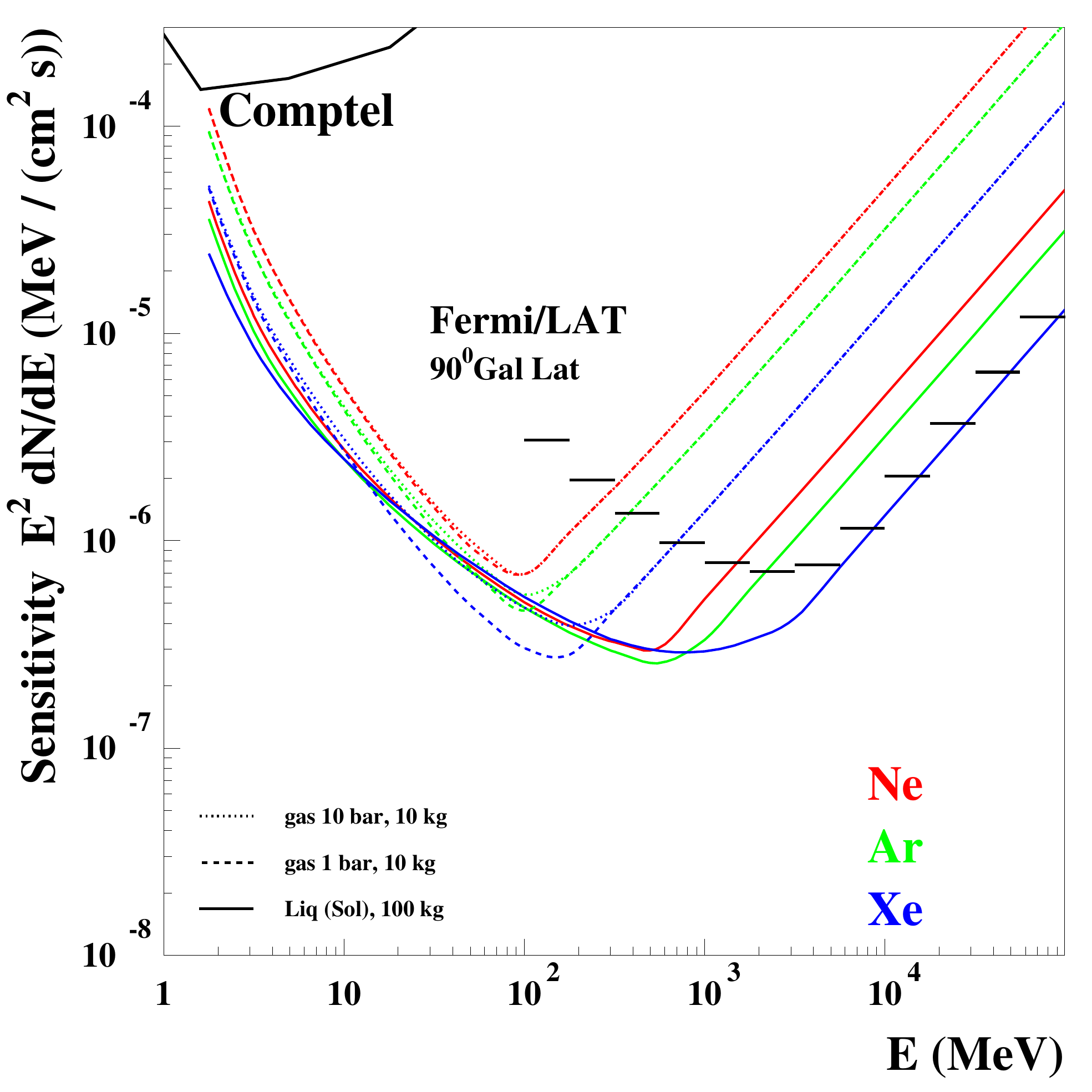}
\caption{\label{fig:sensitivity}
(Color online).
Dependence of the differential sensitivity on photon energy 
compared to the $90^\circ$ galactic latitude performance of the Fermi/LAT
\cite{Fermi:1206.1896} and of the Compton telescope 
COMPTEL \cite{Schönfelder}.
Left : the ideal case in which the missing recoil momentum would be
neglected in the expression of $\sigma_\theta $.
Right : the real case, with the full expression of $\sigma_\theta $.
}
\end{center}
\end{figure*}

\subsection{Poisson statistics}

Using Poisson statistics, and following the method used by
Ref. \cite{Abdo:2010ru}
(A recent derivation of eq. (\ref{eq:significance:poisson}) can be
found in Ref. \cite{Cowan:2010js} eq. (96)),
 the significance $n$ becomes such that:
\begin{equation}
n^2 = 2 B [(1+S/B) \ln(1+S/B) - S/B], 
\label{eq:significance:poisson}
\end{equation}
which reduces to $n^2 = S^2/B$ 
in the background-dominated regime
 $S/B \ll 1$, as expected.
We solve eq. (\ref{eq:significance:poisson}) for $S$, with $n=5$.

Figure \ref{fig:sensitivity} shows the dependence on energy  of the
differential sensitivity for the nine test cases of this study 
compared to that of the Fermi/LAT \cite{Fermi:1206.1896} and of the
Compton telescope COMPTEL \cite{Schönfelder}. 
On the left plot, an ideal case for which the missing recoil momentum
could be neglected in the expression of the angular resolution, we see
that a gas TPC  would clearly have a better performance than a liquid TPC:
the better angular resolution would outweigh  the smaller
sensitive mass (10 kg compared to 100 kg).
In reality, the effect of the missing recoil momentum on the
sensitivity is visible at low energy (right plot).

Together with Compton telescope projects, which
 aim at sensitivities of the order of $10^{-5}
\mega\electronvolt/(\centi\meter^2 \second$) close to 1 MeV \cite{Diehl:2011qg}, there
is good hope to fill the sensitivity gap between the energy ranges of the
Compton and pair telescopes, and even to envisage enough overlap for
a cross-calibration.

\section{Track momentum from multiple measurement of multiple scattering}

In this section we examine the potential of a TPC to determine the
momentum of the conversion electrons from the deflection due to
multiple scattering: since the average deflection angle $\theta_0$ is
proportional to $1/p$, each deflection provides a momentum measurement
\cite{Pinkau:1972sx}.
Bolton describes the relative momentum resolution $\sigma_p/p$ as
asymptotes of a ``universal function'' \cite{Bolton:1997rm}: 
\begin{itemize}
\item
In the multiple scattering dominated regime (low momentum):
$
\gfrac{\sigma_p}{p} = \frac{1}{\sqrt{2N}},
$
where $N$ is the number of samplings.

\item
In the spatial resolution dominated regime (high momentum):
$
\gfrac{\sigma_p}{p} = \frac{1}{\sqrt{2N}} 
\left( \frac{p}{p_m} \right)^2,
$
where $p_m = \gfrac{p_0 \Delta^{3/2}}{\sigma \sqrt{X_0}}$, and
$\Delta$ is the length over which the deflection is measured, which
must be larger than the sampling of the TPC, $\Delta > l$.
\end{itemize}

Approximating the full expression by the sum of the two asymptotes, 
we obtain:
\begin{equation}
\frac{\sigma_p}{p} = \frac{1}{\sqrt{2L}} 
\left[
\Delta^{1/2}
+
\frac{p^2 \sigma^2 X_0}{\Delta^{5/2} p_0^2} 
\right], 
\label{eq:deux}
\end{equation}

the minimum of which is obtained for:
\begin{equation}
 \Delta = 
\left[
\frac{5 p^2 \sigma^2 X_0}{ p_0^2}
\right]^{1/3}.
\label{eq:Delta}
\end{equation}

\begin{figure} [th]
 \begin{center} 
\includegraphics[width=0.864\linewidth]{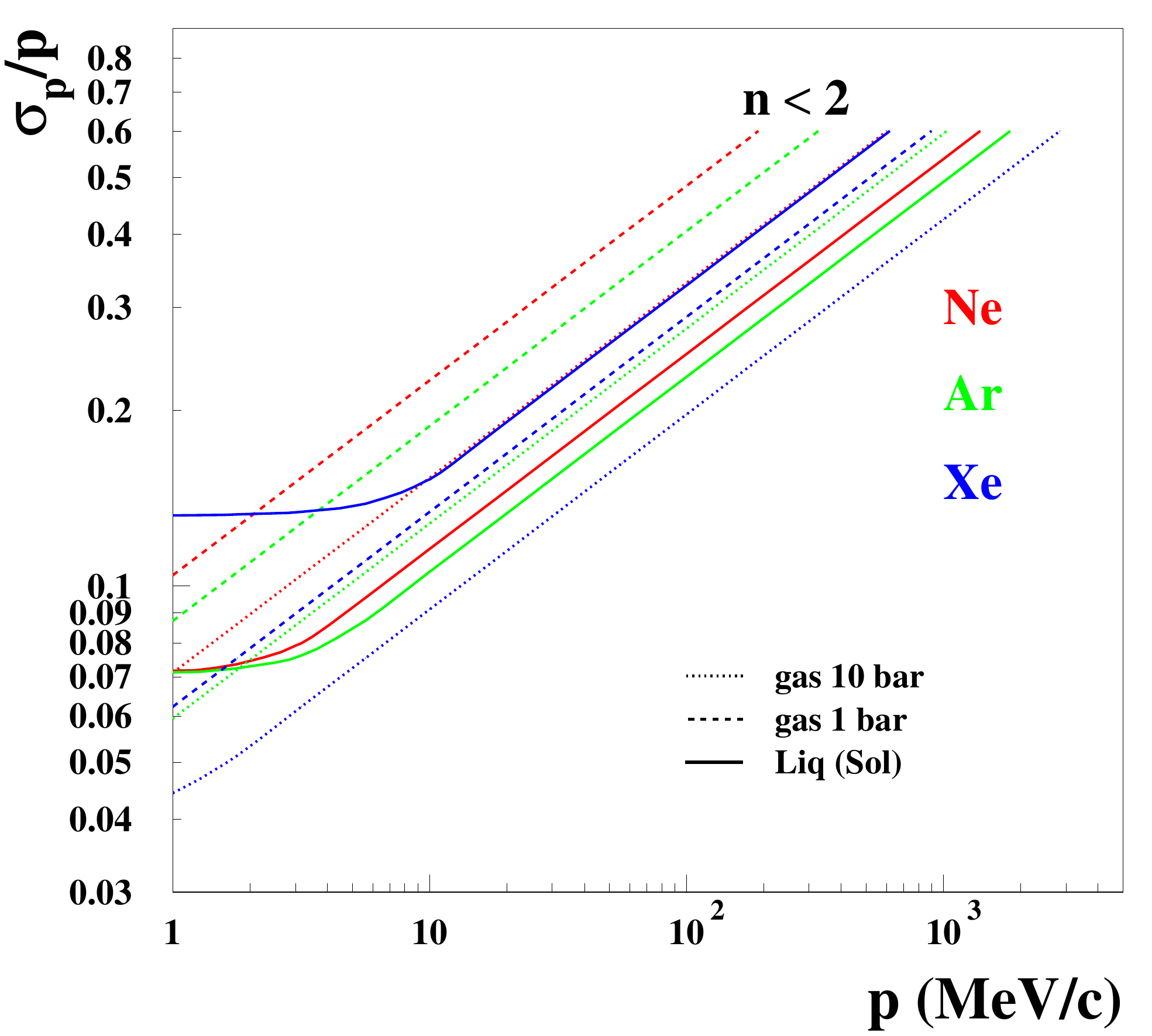}
\caption{\label{fig:mmms}
(Color online).
Relative track momentum resolution for optimal sampling $\Delta$ as a function of track momentum.
(The low momentum asymptote for the liquids is due to the $\Delta > l$ limit).
}
\end{center}
\end{figure}

As the momentum is not known a priori, some iteration will be needed.
The value of $\sigma_p/p$ for that optimal set is: 
\begin{equation}
\frac{\sigma_p}{p} = 
\frac{C}{\sqrt{2L}} 
\left[ \frac{p}{p_0} \right]^{1/3}
\left[ \sigma^2 X_0 \right]^{1/6}
\label{eq:resolution:min}
\end{equation}

and $C \equiv 5^{1/6} + 5^{-5/6} \approx 1.57 $.
The method is usable at low energy, below 100 MeV (Fig. \ref{fig:mmms}.)

\section{Summary}

We examined the performance of a TPC as a $\gamma$-ray telescope
above the pair-creation threshold.
Analytical expressions are obtained for the various contributions to
the photon angular resolution and the track momentum resolution.
These analytical results are exemplified numerically.
Even with the modest value of the parameters used in the present study, 
a $L = 30 \, \centi\meter$ tracking length and $M = 10\, \kilo\gram$
sensitive mass for gases and
 $L = 10 \, \centi\meter$ tracking length and $M = 100\, \kilo\gram$ for
liquids, TPCs show  huge potential in the MeV-GeV range.

The angular resolution is found to have an $E^{-3/4}$ energy
dependence in the multiple scattering dominated energy range, except
for gas TPC at low energy, where the missing recoil momentum dominates
with a steeper dependence close to $E^{-5/4}$.
An improvement of a factor better than ten in angular resolution with
respect to the present tungsten-slab-converter telescopes is within
reach for a gas TPC.
This provides at least two orders of magnitude in background
rejection, not to mention the easy albedo rejection and the immunity
to cosmic-ray pile-up thanks to the detailed imaging provided by a
TPC.
The improved angular resolution translates also into a dramatic
improvement of the differential sensitivity that will bridge up with a
future Compton mission at a couple of MeV at a level of about $10^{-5}
\mega\electronvolt/(\centi\meter^2 \second)$.

Track momentum can be measured from the deflections due to
multiple scattering across the detector.
After optimization of the sampling size, 
the momentum relative precision is found to vary like $p^{1/3}$.
The method can be used for momenta smaller than 
$100 \, \mega\electronvolt/c$.
Above that limit, the photon energy measurement must be performed by an
additional device.
A TPC might well be a first layer of an otherwise high-energy 
(0.1 GeV - 3 TeV) range project such as Gamma-400
\cite{Galper:2012fp}.

An R\&D program is in progress to characterize the use of a TPC as a
$\gamma$-ray telescope and as a $\gamma$-ray polarimeter
\cite{Bernard:2012em,Bernard:2012jy}.

A number of approximations have been made in this work so as to
obtain scaling laws in the form of analytical expressions that are
easily handled.  
To name a few, we have neglected the logarithmic correction of multiple
scattering, the loss of energy of the tracks inside the detector,
their emission of radiation that disturbs tracking at high momentum, 
especially for high-$Z$ detectors and/or dense TPC -- such as liquid
xenon, and trigger inefficiency.
These limitations would be addressed using a full simulation, 
which is beyond the scope of the present study.


\begin{thebibliography}{99}
\bibitem{Ghisellini:2012hs} 
 G.~Ghisellini,
 ``Radiative Processes in High Energy Astrophysics,''
 arXiv:1202.5949 [astro-ph.HE].


\bibitem{Hartman}
Hartman, R. C.,
``Astronomical gamma ray telescopes in the pair production regime'',
Nucl Physics B Proc. Suppl., 10 (1989) 130.


\bibitem{Hunter:2001}
Hunter, S. D.; Bertsch, D. L.; Deines-Jones, P.
``Design of a Next Generation High-Energy Gamma-ray telescope'',
GAMMA 2001: Gamma-Ray Astrophysics 2001. AIP Conference Proceedings, 587, 848-852 (2001).

\bibitem{Ueno:2011}
K. Ueno {\it et al.},
``Development of the tracking Compton/pair-creation camera based on a gaseous TPC and a scintillation camera'',
Nucl.\ Instrum.\ Meth.\ A {\bf 628}, 158 (2011).

\bibitem{Nygren:1978}
D.R. Nygren and J. N. Marx,
``The Time Projection Chamber'', 
Physics Today 31 (1978) 46.

\bibitem{Attie:2009zz} 
 D.~Attie,
 ``TPC review,''
 Nucl.\ Instrum.\ Meth.\ A {\bf 598}, 89 (2009).

\bibitem{Bondar:2005wx} 
 A.~Bondar
 ``Two-phase argon and xenon avalanche detectors based on gas electron multipliers,''
 Nucl.\ Instrum.\ Meth.\ A {\bf 556}, 273 (2006).

\bibitem{Badertscher:2010zg}
 A.~Badertscher {\it et al.},
 ``First operation of a double phase LAr Large Electron Multiplier Time
Projection Chamber with a two-dimensional projective readout anode,''
 Nucl.\ Instrum.\ Meth.\ A {\bf 641} (2011) 48.

 
\bibitem{Brisson:1983qb} 
 V.~Brisson {\it et al.},
 ``Performance Of A Prototype Solid Neon And Solid Argon Calorimeter,''
 Nucl.\ Instrum.\ Meth.\ {\bf 215}, 79 (1983).

\bibitem{Gluckstern} 
R.~L.~Gluckstern, 
``Uncertainties in track momentum and direction, due to multiple scattering and measurement errors,'' 
Nucl.\ Instrum.\ Meth.\ {\bf 24}, 381 (1963).

\bibitem{Innes:1992ge}
W.~R.~Innes,
``Some formulas for estimating tracking errors,''
 Nucl.\ Instrum.\ Meth.\ A {\bf 329}, 238 (1993).

\bibitem{PDG}
J. Beringer {\it et al.},
(Particle Data Group), Phys. Rev. D86, 010001 (2012). 

\bibitem{Fermi:1206.1896}
``The Fermi Large Area Telescope On Orbit: Event Classification, Instrument Response Functions, and Calibration'',
Fermi-LAT Collaboration,
arXiv:1206.1896

\bibitem{Jost:1950zz} 
 R.~Jost, J.~M.~Luttinger and M.~Slotnick,
 ``Distribution of Recoil Nucleus in Pair Production by Photons,''
 Phys.\ Rev.\ {\bf 80}, 189 (1950).

\bibitem{Heitler1954}
``The quantum theory of radiation'',
W. Heitler,
1954.

\bibitem{Borsellino1953}
``Momentum Transfer and Angle of Divergence of Pairs Produced by Photons'',
 A. Borsellino, 
Phys. Rev. 89, 1023 - 1025 (1953).

\bibitem{Mott:1934}
N.F. Mott, H.S.W. Massey, 
``The Theory of Atomic Collisions'', University Press, Oxford, 1934.

\bibitem{Olsen:1963zz} 
 H.~Olsen,
``Opening Angles of Electron-Positron Pairs,''
 Phys.\ Rev.\ {\bf 131}, 406 (1963).

\bibitem{Aprile:2008ft} 
 E.~Aprile {\it et al.},
 ``Compton Imaging of MeV Gamma-Rays with the Liquid Xenon Gamma-Ray Imaging Telescope (LXeGRIT),''
 Nucl.\ Instrum.\ Meth.\ A {\bf 593}, 414 (2008)
 [arXiv:0805.0290 [physics.ins-det]].

\bibitem{NIST}
NIST, National Institute of Standards and Technology,
Physical Reference Data,
http://physics.nist.gov/PhysRefData/

\bibitem{Abdo:2010nz} 
 A.~A.~Abdo {\it et al.} [Fermi-LAT Collaboration],
 ``The Spectrum of the Isotropic Diffuse Gamma-Ray Emission Derived From First-Year Fermi Large Area Telescope Data,''
 Phys.\ Rev.\ Lett.\ {\bf 104}, 101101 (2010)
 [arXiv:1002.3603 [astro-ph.HE]].

\bibitem{Abdo:2010ru} 
 A.~A.~Abdo {\it et al.} [Fermi-LAT Collaboration],
 ``Fermi Large Area Telescope First Source Catalog,''
 Astrophys.\ J.\ Suppl.\ {\bf 188}, 405 (2010)
 [arXiv:1002.2280 [astro-ph.HE]].

\bibitem{Cowan:2010js} 
 G.~Cowan, K.~Cranmer, E.~Gross and O.~Vitells,
``Asymptotic formulae for likelihood-based tests of new physics,''
 Eur.\ Phys.\ J.\ C {\bf 71}, 1554 (2011).

\bibitem{Schönfelder}
``Lessons learnt from COMPTEL for future telescopes'',
Volker Schönfelder,
New Astronomy Reviews,
 48 2004 193.

\bibitem{Diehl:2011qg} 
 R.~Diehl,
 ``GRIPS and the Perspective of Next-generation Gamma-ray Surveys,''
 PoS INTEGRAL {\bf 2010}, 035 (2010)
 [arXiv:1107.4892 [astro-ph.IM]].

\bibitem{Pinkau:1972sx} 
 K.~Pinkau,
 ``Analysis procedure of gamma ray astronomy spark chamber data,''
 Nucl.\ Instrum.\ Meth.\ {\bf 104}, 517 (1972).

\bibitem{Bolton:1997rm} 
 T.~Bolton,
 ``High-energy muon momentum estimation from multiple Coulomb scattering in dense detectors,''
 hep-ex/9705007.

\bibitem{Galper:2012fp} 
 A.~M.~Galper {\it et al.},
 ``Status of the GAMMA-400 Project,''
 arXiv:1201.2490 [astro-ph.IM].

\bibitem{Bernard:2012em} 
 D.~Bernard and A.~Delbart,
``High-angular-precision gamma-ray astronomy and polarimetry,''
 Nucl.\ Instrum.\ Meth.\ A {\bf 695}, 71 (2012).


\bibitem{Bernard:2012jy} 
 D.~Bernard,
 ``HARPO - A Gaseous TPC for High Angular Resolution Gamma-Ray Astronomy and Polarimetry from the MeV to the TeV,''

 Nucl.\ Instrum.\ Meth.\ A, In Press,
 arXiv:1210.4399 [astro-ph.IM].

\end{thebibliography}
\end{document}